\def\L{{\cal L}}
\documentstyle[12pt]{article}
\begin{document}
\begin{flushright}
to appear in Phys.~Lett.~B \\
(submitted in November 1992)
\end{flushright}
\centerline{\LARGE \bf CP-Violation with Bosons}
\vspace{2.cm}
\begin{center}
{{\large G.~Cveti\v c, M.~Nowakowski} \\
Inst.~f\"ur Physik, Universit\"at Dortmund, 4600 Dortmund 50, Germany \\[0.8cm]
{\large A.~Pilaftsis} \\
Inst.~f\"ur Physik, Johannes-Gutenberg Universit\"at, 6500 Mainz, Germany}
\end{center}
\vspace{2.cm}
\centerline{\bf ABSTRACT}

We formulate the conditions under which a purely bosonic theory (without
fermions) containing neutral spin-0 particles and vector (gauge) bosons
violates the CP-symmetry through the presence of CP-even and CP-odd
operators in the Lagrangian. This is done without reference to
explicitly CP-violating scalar sector of extended standard models (SM)
with two or more Higgs doublets. The Lagrangian expressed in the mass
basis of the spin-0 fields can, however, in certain cirumstances be
identified with a part of the CP-violating SM with two Higgs doublets.
It is instructive to consider the manifestation of CP-violation in the
mass basis since this leads directly to suggestions of genuine CP-signals.

\newpage
\vspace{1cm}
CP-violation has always been a subject of numerous investigations in
particle physics ~\cite{pt}.
One of the reasons is that CP-violation is an
important ingredient in explaining the baryon asymmetry of the universe
{}~\cite{as}.
It is known that the strength of the CP-violation coming from the
Kobayashi-Maskawa sector of the standard model (SM) is not sufficient
to explain this asymmetry ~\cite{ckn}.
It is therefore important to consider also other sources of CP-violation.
Recently, the CP-violating SM with two Higgs doublets has attracted
a lot of attention ~\cite{wm}
since it could give a natural explanation of the baryogenesis ~\cite{ckn,tz}.

Usually, such a theory is formulated in the isoweak basis, i.e.~with
2$SU(2)_L$-doublets. However, since the physical interaction terms are
given in the mass basis of the particles, it is very instructive (and
closer to the experimental situation) to view the manifestation of
CP-violation from this end (mass basis). Later on, we can choose to
identify the bosonic part of the Lagrangian with that of the extended
SM. Our considerations of CP-violation are, in principle, independent of
any specific choice of electroweak theory. From this more pragmatic
point of view, CP-violation in the mass basis of the theory is manifested
through the simultaneous presence of CP-even and CP-odd terms in the
Lagrangian. More precisely, we cannot assign CP-eigenvalues to the
mass eigenstates in such a way that the Lagrangian is CP-invariant.
However, the coupling parameters are real.
In our considerations, we will ignore fermions.

First, let us consider two well-known examples of Lagrangians in which
at least two terms with different CP-transformation properties
simultaneously appear to give a CP-violating theory.

One such example is the following interaction of a neutral spin-0 boson
$\varphi^o$ with fermions ($f \bar f$)
\begin{equation}
\L_{\varphi^o f \bar f} = \varphi^o \lbrace \alpha \bar \Psi \Psi
+ i \beta \bar \Psi \gamma_5 \Psi \rbrace  \qquad
( \alpha, \beta \in {\rm R}).
\end{equation}
It is known that the Lagrangian (1) is part of a spontaneously broken
theory ~\cite{l}
where the vacuum is not an eigenstate of CP. The Lagrangian of this theory,
as given in terms of the non-shifted field $\phi^o = v + \varphi^o$ ($v$
is the vacuum expectation value), is CP-invariant
if we assign $CP(\phi^o) = -1$.
It is then clear that $\varphi^o$ is not a CP-eigenstate.
However, the fact that $\L_{\varphi^o f \bar f}$ is CP-nonconserving is
independent of these involved considerations, i.~e.~independent of the
assumption of the spontaneous symmetry breaking. Namely, we can simply
look at (1) as a given interaction Lagrangian of physical fields. It
is then legitimate to perform a CP-transformation of the (neutral)
physical field $\varphi^o$ and to try to assign $\pm 1$ CP quantum numbers
to it. Since the Lagrangian (1) contains a linear combination of
CP-even ($CP(\bar \Psi \Psi) = 1$) and CP-odd ($CP(i \bar \Psi
\gamma_5 \Psi) = -1$) fermionic operators, it is impossible to assign
any CP-eigenvalue to $\varphi^o$ such that $\L_{\varphi^o f \bar f}$
is CP-invariant.

The other example is a purely bosonic Lagrangian which involves only
spin-1 physical fields. Let $V_{\mu}$ and $W_{\mu}$ be neutral and
charged vector fields, respectively. If we restrict ourselves to
dimesion-4 operators, we can construct the following interaction terms:
\begin{eqnarray}
\L^{(1)}_{spin-1} & = & i \kappa^{(1)} [ W^{-}_{\mu \nu}
W^{+ \mu}V^{\nu} - W^{+}_{\mu \nu}V^{\nu}W^{- \mu} ] +
  \nonumber\\
 & & + i \kappa^{(2)} W^{-}_{\mu}W^{+ \nu}V^{\mu \nu},
\end{eqnarray}
\begin{equation}
\L^{(2)}_{spin-1}  =  \kappa^{(3)} W^{+}_{\mu}W^{-}_{\nu}
 [\partial^{\mu}V^{\nu} + \partial^{\nu}V^{\mu}],
\end{equation}
where
\begin{eqnarray}
W_{\mu \nu} & = & \partial_{\mu}W_{\nu} - \partial_{\nu}W_{\mu},
 \nonumber\\
V_{\mu \nu} & = & \partial_{\mu}V_{\nu} - \partial_{\nu}V_{\mu}.
\end{eqnarray}
If we set $\kappa^{(1)} = \kappa^{(2)} = g \cos \theta_W$, then
$\L^{(1)}_{spin-1}$ describes the gauge interaction of $Z$ with
charged W-bosons in SM.
The requirement of CP-invariance would now dictate $J^{PC}(V) = 1^{--}$ for
$\L^{(1)}_{spin-1}$ and $J^{PC}(V) = 1^{-+}$ for $\L^{(2)}_{spin-1}$
\footnote{We mention here that the quantum numbers $1^{-+}$ are exotic
in a sense that such a particle cannot couple to fermion-antifermion pair
if parity (P) and C are conserved ~\cite{gn}.}.
Therefore, as before, the linear combination $\L = \L^{(1)}_{spin-1} +
\L^{(2)}_{spin-1}$ violates CP. Note that, if we insist on having
$\L_{spin-1}$ as a part of an $SU(2) \otimes U(1)$ gauge theory, then
$\kappa^{(3)} = 0$ and the neutral vector boson is (in the absence of
other, possibly CP-violating interactions) a $1^{--}$ boson.

We now raise the question: under what circumstances would the spin-0 neutral
sector lead to CP-violation? The latter seems impossible for Lagrangians
which contain {\it only} spin-0 neutral fields. Therefore, the situation
here is more involved than in the case of the Lagrangians (1), or (2) and
(3). Additionally, we have to include at least spin-1 bosons of SM in the
Lagrangian to get such effects.

Let us start with the Lagrangian $(V=Z)$
\begin{eqnarray}
\L_{H^o_i} & = & \L_{kin} + \L_{mass} + g_{ijk}H^o_iH^o_jH^o_k +
 \nonumber\\
 & & + g_{ij}(H^o_i \stackrel{\leftrightarrow}{\partial}_{\mu}H^o_j)
  Z^{\mu} +
 \nonumber\\
 & & + g^Z_iH^o_iZ_{\mu}Z^{\mu} + g^W_iH^o_iW^{+}_{\mu}W^{- \mu},
\end{eqnarray}
where $H^o_i$ are $N$ spin-0 neutral fields which correspond to mass
eigenstates, the indices $i,j,k$ run over $1,\ldots,N$ such that
$i \leq j \leq k$,
and $\stackrel{\leftrightarrow}{\partial}$ is the antisymmetrized derivative:

\begin{displaymath}
H^o_i \stackrel{\leftrightarrow}{\partial}_{\mu}H^o_j =
H^o_i \vec \partial_{\mu} H^o_j - H^o_j \vec \partial_{\mu} H^o_i.
\end{displaymath}
Note that in SM with one or two Higgs doublets, the terms containing
physical neutral spin-0 particles (to up to 3rd power) represent
just special cases of the Lagrangian (5).
If we assume first that {\it all} coupling parameters $g_{ijk}$ and
at least one parameter $g_{ij}$ in (5) are nonzero, then
$\L_{H^o_i}$ is invariant under CP only if $J^{PC}(H^o_i) = 0^{++}$ and
$J^{PC}(Z) = 1^{-+}$.
\footnote{Note that the assigned eigenvalues of P
yield P-invariant $\L_{H^o_i}$ {\it and} $\L^{(j)}_{spin-1}$ (j=1,2),
which implies that the question of CP-violation reduces in our cases to
the question of C-violation.}
More precisely, $\L_{H^o_i}$ is invariant
under the transformations
\begin{eqnarray}
PH^o_i(t, \vec x )P^{-1} & = & \eta^P(H^o_i)H^o_i(t,- \vec x ), \
\eta^P(H^o_i)=1,
 \nonumber\\
CH^o_i(x)C^{-1} & = & \eta^C(H^o_i)H^o_i(x), \
\eta^C(H^o_i)=1,
 \nonumber\\
PZ^{\mu}(t, \vec x )P^{-1} & = & \eta^P(Z)Z_{\mu}(t,- \vec x ), \
\eta^P(Z)=1,
 \nonumber\\
CZ^{\mu}(x)C^{-1} & = & \eta^C(Z)Z^{\mu}(x), \
\eta^C(Z)=1.
\end{eqnarray}
The Lagrangian
\begin{equation}
\L^{(2)} = \L_{H^o_i} + \L^{(2)}_{spin-1}
\end{equation}
is then also CP-invariant, but the Lagrangian
\begin{equation}
\L^{(1)} = \L_{H^o_i} + \L^{(1)}_{spin-1}
\end{equation}
violates CP (it is P-invariant, but violates C). The reason is that we
cannot assign any CP quantum numbers to the fields to make $\L^{(1)}$
CP-invariant. At least two neutral spin-0 particles must be
present in order to violate CP in $\L^{(1)}$ in such a case
(note that SM with two Higgs
doublets contains three neutral Higgses). We can then indeed
interpret the $H^o_i$ fields as linear combinations of CP-even and
CP-odd (non-physical) fields. Note that it is irrelevant whether this
mixing takes place on a more fundamental level, e.~g.~in a two Higgs
SU(2)-doublet model, or if (8) is an effective Lagrangian of composite
fields $H^o_i$.

The Lagrangian (8) is not the only CP-violating combination. Typical
charged scalar field interaction of the form
\begin{equation}
\L_{ZH^{+}H^{-}} = i g_{ZH^{+}H^{-}}(H^{+}
\stackrel{\leftrightarrow}{\partial}_{\mu} H^{-}) Z^{\mu}
\end{equation}
is CP-invariant only if again $J^{PC}(Z) = 1^{--}$. Therefore, the same
arguments as before lead us to the statement that the Lagrangians
\begin{eqnarray}
\L_{H^oH^{\pm}} & = & \L_{H^o_i} + \L_{ZH^{+}H^{-}},
 \nonumber\\
\L^{(3)} & = & \L^{(1)}_{spin-1} + \L_{H^o_i} + \L_{ZH^{+}H^{-}}
\end{eqnarray}
also violate the CP-symmetry.

These conclusions were arrived at when assuming that all coupling parameters
$g_{ijk}$ and at least one $g_{ij}$
in $\L_{H^o_i}$ (eq.(5)) are nonzero. On the other hand, it is possible
to restore CP-invariance in $\L^{(3)}$ (or $\L^{(1)}$) by putting some
coupling parameters to zero. For instance, if there are
just {\it two} neutral $H^o_i$ fields (i=1,2),
then we restore it if, for example
\begin{equation}
g_{211} = g_{222} = g_2^Z = g_2^W = 0
\end{equation}
- by assigning $J^{PC}(H^o_1) = O^{++}$,
$J^{PC}(H^o_2) = O^{+-}$ (exotic quantum numbers).

Similarly, for {\it three} physical neutral $H^o_i$-fields, CP in
$\L^{(3)}$ (or $\L^{(1)}$) is restored if:
\begin{enumerate}
\item There is at least one specific $H^o_i$ (say $H^o_1$) which appears in
all nonzero $(H^o_jH^o_kH^o_l)$-terms as an {\it odd} power (as $(H^o_1)^1$
and/or $(H^o_1)^3$) and the corresponding other couplings are zero:
\begin{equation}
g_{222}=g_{333}=g_{112}=g_{113}=g_{23}=g_2^Z=g_3^Z=g_2^W=g_3^W=0.
\end{equation}
(Possible assignments are: $J^{PC}(H^o_1) = O^{++}$,
$J^{PC}(H^o_2)=J^{PC}(H^o_3)=O^{+-}$.)

\item There is at least one specific $H^o_i$ (say $H^o_1$) which appears in
all nonzero $(H^o_jH^o_kH^o_l)$-terms as an {\it even} power (as
$(H^o_1)^2$ and/or zeroth power)
and the corresponding other couplings are zero:
\begin{equation}
g_{122}=g_{133}=g_{123}=g_{111}=g_{23}=g_1^Z=g_1^W=0.
\end{equation}
(Possible assignments are: $J^{PC}(H^o_1) = O^{+-}$,
$J^{PC}(H^o_2)=J^{PC}(H^o_3)=O^{++}$.) One such specific case is SM with
two Higgs doublets and no CP-violation ($\xi = 0$, where
$<0 \mid \Phi_2^o \mid 0> = v_2 e^{i \xi}$, in the notation
of ref. ~\cite{gh}).
\end{enumerate}

We necessarily obtain CP-violation in $\L^{(3)}$ (or $\L^{(1)}$),
if at least one of the terms
$(H^o_i \stackrel{\leftrightarrow}{\partial}_{\mu} H^o_j)
Z^{\mu}$ is nonzero {\it and}
the terms ($H^o_jH^o_kH^o_l$) satisfy
neither 1.~nor 2.~(i.~e.~, each of the three $H^o_i$'s appears in
$H^o_jH^o_kH^o_l$-terms at least once as an odd power and at least once as
an even power) - for example, if all the nonzero $g_{jkl}$ are:
$g_{123},g_{112},g_{223} \ \mbox{and} \ g_{233}$.
CP-invariance of the $(H^o_jH^o_kH^o_l)$-terms alone would imply that
$J^{CP}(H^o_i) = O^{++}$, and similarly for $\L^{(1)}_{spin-1}$
$J^{CP}(Z) = 1^{--}$ - but then the appearance of the nonzero coupling
$(H^o_i \stackrel{\leftrightarrow}{\partial}_{\mu} H^o_j)Z^{\mu}$
violates CP.
Note that SM with two Higgs doublets
and $\xi \not= 0$ (more precisely: $(\lambda_5 - \lambda_6) \sin (2 \xi )
\not= 0$, in the notation of ref.~\cite{gh})
turns out to be one such case, as can be checked explicitly
by diagonalizing the Higgs sector ~\cite{cn}.

It is worth noticing that in general $\L_{H^o_i} + \L^{(1)}_{spin-1}$
or $\L_{H^o_i} + \L_{ZH^+H^-}$, with all couplings $g_{ijk}$
and at least one $g_{ij}$ being nonzero and
the number of spin-0 particles $N \geq 2$, suffice to establish
CP-violation in these models even though we may have additional interaction
terms like quartic couplings, etc.

The mass basis in which we have written the Lagrangians has the advantage
that we can now look directly for processes that would reveal CP-violation
in the bosonic sector of the Lagrangian. In general, this could be realized
directly, by predicting and observing asymmetries of the form
\begin{equation}
\bigtriangleup_{AB} = dN( \mid A \rangle \rightarrow \mid B \rangle ) -
 dN( \mid \bar A \rangle \rightarrow \mid \bar B \rangle ),
\end{equation}
where $\mid \bar A \rangle$ and $\mid \bar B \rangle$
are CP-conjugated states and
$dN$ is the number of events. On the other hand, CP-violation could be
observed (and predicted) also indirectly, by two nonzero amplitudes of
the form
\begin{eqnarray}
T( \mid A \rangle \rightarrow \mid B \rangle_{CP=+1})
 \nonumber \\
T( \mid A \rangle \rightarrow \mid B \rangle_{CP=-1}).
\end{eqnarray}
The possibility (15) implies that one should construct CP-even and
CP-odd states (components). Feynman rules are usually formulated in the
basis of spin or helicity eigenstates which are not eigenstates of CP.
This, of course, does not mean that one cannot perform discrete symmetry
tests in the helicity basis; but we prefer to do it in the basis of
orbital angular momentum eigenstates which have the advantage of being also
CP-eigenstates (see below).

Restricting ourselves to tree level processes (i.~e.~, we do not take into
account higher derivative couplings like $H^o_iW^-_{\mu \nu}W^{+ \mu \nu}$,
$H^o_iZ^{\mu \nu}Z_{\mu \nu}$, which can be effectively generated at
one-loop level), we observe that the two-particle final states of decay
processes, as predicted by (CP-violating) Lagrangian $\L^{(3)}$ (eq.~(10)),
are either S-wave,
or P-wave (derivative coupling). Taking in (15) $\mid A \rangle$ as the
spin-0 particle state
($\vec J = \vec L + \vec S = \vec 0$) and $\mid B \rangle$ as
a particle-antiparticle state, then the latter state when produced at tree
level is an S-wave ($\vec L_{final} = \vec 0 = \vec S_{final}$) and cannot
contain both components of CP (since $CP(B) = (-1)^{S_{fin}} = +1$ for
boson-antiboson final state). If we want to have a mixture of both
CP-components, we have to consider $1 \rightarrow 3$ processes like
\begin{eqnarray}
H^o_i & \rightarrow & W^+W^-Z,
 \nonumber\\
H^o_i & \rightarrow & H^+H^-Z,
\end{eqnarray}
These processes are the simplest examples which would give a genuine
signal of CP-violation in the neutral spin-0 sector (without the inclusion
of fermions). Of course, we assume that the mass spectrum of the theory
allows kinematically these decays. A correct procedure to show that the
final states of these reactions are mixtures of $CP=1$ and $CP=-1$
components would be to perform a partial wave analysis of the amplitudes
in terms of orbital and spin angular momenta for 3-particle final state.
This basis has the properties
\begin{eqnarray}
P \mid l \sigma ; L \Sigma \rangle & = & \eta^P_3 (-1)^{l+L}
\mid l  \sigma; L \Sigma \rangle,
 \nonumber\\
C \mid l \sigma ; L \Sigma \rangle & = & \eta^C_3 (-1)^{l+ \sigma}
\mid l \sigma; L \Sigma \rangle,
\end{eqnarray}
where $l$ and $\sigma$ are the relative orbital angular momentum and
the resultant spin of the particle-antiparticle pair (1-2), $L$ and
$\Sigma$ are the corresponding quantum numbers of the neutral
particle 3 and the subsystem (1-2) as a whole ~\cite{w}.
The phases $\eta_3$'s are the intrinsic quantum numbers of the neutral
particle 3 (which we will take to be $Z$, i.~e.~ the final state is
$W^+W^-Z$). In a CP-violating theory, $\eta_3$'s are, in principle,
conventional. Through the detected angular distribution
$dN/d \Omega$,
it should be possible to show whether or not both CP components
$(W^+W^-Z)_{CP=1}$ and $(W^+W^-Z)_{CP=-1}$ are produced in the decay
process. We plan to do such an analysis in the near future, in the case
of SM with two Higgs doublets ($\xi \not= 0$ ~\cite{gh}).
Here, we will give another
argument for the presence of CP-even and CP-odd final states in (16).
It suffices to show this for the first of these two processes ~\cite{ll}.
The line of arguments for the other is similar.

There are several amplitudes contributing to $H^o_i \rightarrow W^+W^-Z$
if the underlying dynamics is described by $\L^{(1)}$ of eq.~(8). The
following three are possible also in the minimal SM or in a CP-conserving
version of SM with two Higgs doublets:
\begin{eqnarray}
T(H^o_i \rightarrow W^+W^{- \star} \rightarrow W^+W^-Z),
 \nonumber\\
T(H^o_i \rightarrow W^-W^{+ \star} \rightarrow W^-W^+Z),
 \nonumber\\
T(H^o_i \rightarrow Z^{\star}Z \rightarrow W^+W^-Z).
\end{eqnarray}
The star in (18) denotes an off-shell intermediate particle state. It
follows that only states with $CP(W^+W^-Z) =+1$ contribute to the
diagrams of (18).
The crucial point is now that the CP-properties of the final state
(as given in (17)) are determined only by the Lorentz structure of
the interaction terms (when adopting the convention that the internal
$CP= \eta^C_3 \eta^P_3$ of the final particles is $+1$)
and are independent of the magnitude of the
(real) coupling parameters. The conclusion that $CP(W^+W^-Z)=+1$ in
(18) is then the same for a CP-conserving and CP-violating theory
(although the initial state is a linear combination of both
CP-eigenstates with $CP=+1$ and $CP=-1$ in a CP-violating theory).
However, a fourth amplitude contributing to the process in
a CP-violating theory does not exist
in the minimal SM, namely
\begin{equation}
\sum_j T(H^o_k \rightarrow H_j^{o \star}Z \rightarrow W^+W^-Z).
\end{equation}
This amplitude contains a momentum-dependent vertex $(H^o_k
\stackrel{\leftrightarrow}{\partial}_{\mu} H^o_j) Z^{\mu}$
and yields final states with only the $CP= -1$ component (both types of
amplitudes, (18) and (19), yield nonzero contributions if CP is
violated) . The reason is that the diagram corresponding to (19)
exists also
in the CP-conserving SM with two Higgs doublets ($\xi = 0$), with $H^o_k$
being the "pseudoscalar" ($J^{PC}(H^o_k) = 0^{+-}$). Consistently, we
have in (19) only $CP=-1$ component in the initial and in the final state in
$\xi = 0$ case. In $\xi \not= 0$ case, the initial state in (19) is a linear
combination of two CP-eigenstates with different eigenvalues, and the
final state has only $CP=-1$ (by the same arguments as above for (18)).

{}From the experimental data for the angular distribution
\begin{displaymath}
d \Gamma (H^o_k \rightarrow W^+W^-Z)/ d \Omega_1 d \Omega_2 \ ,
\end{displaymath}
it should in principle be possible to disentangle,
through a partial wave analysis in the orbital angular momentum basis,
the contributions to $\Gamma (H^o_k \rightarrow W^+W^-Z)$
of the components of the final states with $CP=+1$ from those with $CP=-1$
\footnote{Note that states with opposite CP are
orthogonal to each other and hence the interference terms of
amplitudes (18) and (19) yield zero in the decay width
$\Gamma(H^o_k \rightarrow W^+W^-Z)$.}.
It is important to note that such a signal would be, if obtained from
experiments, a genuine signal of CP-violation, i.~e.~a signal independent
of any specific theoretical assumptions.

If we already knew that the underlying theory is SM with two Higgs
doublets, then the evidence of the decays $H^o_i \rightarrow ZZ$
(or $H^o_i \rightarrow W^+W^-$) for all $i=1,2,3$ ~at tree level
would be a signal for CP-violation. However, this would amount to
first ``proving'' the theory in order to prove CP-violation. We regard
as more realistic to deal directly with the processes of eq.~(16).

In conclusion, we state that
we have considered here CP-violation originating from the
bosonic sector with neutral spin-0 particles. Not every such CP-violating
theory can be identified with SM containing two Higgs doublets, since,
in principle, even the existence of just two neutral spin-0 particles
can lead to
CP-violation. We emphasized how CP-violation can manifest itself in the
mass basis of the neutral spin-0 particles, the considerations being
in principle independent of any specific assumptions (SSB, etc.~) on how the
physical, CP-violating interactions came about.
\\
\\[1.cm]
{\bf Acknowledgements.}
We thank E.~A.~Paschos, Y.~L.~Wu and A.~Joshipura for useful discussions
2and comments on the subject. This work has been supported in part by the
German Bundesministerium f\" ur Forschung und Technologie (M.~N., Grant
No.~BMFT 055DO9188) and the Deutsche Forschungsgemeinschaft (G.~C.,
Grant No.~PA254/7-1). The work of A.~P.~ has been supported by a grant
from the Postdoctoral Graduate College of Germany.

\end{document}